\documentclass[12pt,a4paper]{article}
\usepackage{amsmath}
\usepackage{graphicx}
\begin{document}
\begin{titlepage}
\title{ Is elastic scattering at the LHC absorptive or geometric?}
\author{ S.M. Troshin, N.E. Tyurin\\[1ex]
\small  \it Institute for High Energy Physics,\\
\small  \it Protvino, Moscow Region, 142281, Russia}
\normalsize
\date{}
\maketitle

\begin{abstract}
We discuss the role of the impact--parameter dependent quantities. 
Their use along with the available experimental data on elastic scattering at the LHC could serve for an additional insight into the asymptotics. 
\end{abstract}
\end{titlepage}
\setcounter{page}{2}

There is a general opinion that the upper (Froissart-Martin) bound for the total cross-sections 
should be saturated at asymptotics.
The hopes to understand asymptotic behavior of strongly interacting particles always arise  when a  new particle machine
starts to operate. But, after not so long time, it becomes evident again that asymptotic limit is still
elusive. Cosmic rays experiments  probe  much higher energies, but those have much lower statistics. However, their use even allowed to produce a claim that the solution
 of the asymptotic puzzle has been found \cite{blh}.
There are two reasons preventing  acceptance of this conclusion as an only relevant reflection of the physical reality, i.e.
 those results are, in fact,  model-dependent ones and are limited to the use of the
experimental data for the scattering in the forward region only.

Despite that the functional energy dependence of the on-shell hadron total cross-section is considered in most cases to follow $\ln^2 s$-dependence,
the value of the numerical factor in front of $\ln^2 s$ remains obscure. 
This issue is closely
related to the selection of the upper limit for the partial amplitude, namely, should this limit correspond to the
maximum of the inelastic channel contribution to the elastic unitarity with asymptotics
\begin{equation}\label{bd}
  \sigma_{el}(s)/\sigma_{tot}(s)\to 1/2
\end{equation}
or it corresponds to a maximal value of the partial amplidudes allowed by unitarity and results in the asymptotics
\begin{equation}\label{rd}
  \sigma_{el}(s)/\sigma_{tot}(s)\to 1.
\end{equation}

Under assumption of  the limit (\ref{bd}) the original Froissart-Martin bound for the total cross-sections has been improved \cite{mart}. 
The bound for the total inelastic cross-section reduced by factor of 4 has also been derived \cite{mart}.
The ratio  of the elastic to total cross-section (\ref{rd}) corresponds to  energy increase of the total inelastic cross-section slower than  $\ln^2 s$.
It should be noted that this ratio is, in fact, a factor standing in front of $\ln^2 s$ in the bound for the total cross--section\cite{roy}. 
Various asymptotic limits  have been considered in \cite{menon} in almost model-independent way,
but also for the forward scattering data only. 
 
We argue that the inclusion into the analysed data set of the elastic differential cross-section  
can provide new information on the  asymptotics in hadron scattering.
The main tool here is reconstruction of the impact--parameter dependent quantities from this experimental dataset.

 It has been known for a long time (cf. \cite{webb}) that position operator (impact parameter) deprives of
its unpleasant features related to its non-commutativity with the Hamiltonian with  the energy increase.
At the moment the highest available particle collision energy is provided at the LHC.  It seems that at such
high energies  the impact parameter representation for the spinless particles
can be used. One of the attractive features of this representation is  diagonalization of the unitarity equation
for the elastic scattering amplitude $f(s,b)$, i.e.  at high energies
\[
 \mbox{Im} f(s,b)=h_{el}(s,b)+h_{inel}(s,b)
\]
with accuracy 
up to  ${\cal O}(1/s)$ \cite{gold}.
The function $h_{el}(s,b)\equiv |f(s,b)|^2$ is the contribution of the elastic channel, while $h_{inel}(s,b)$ takes into account contribution
of all the  inelastic intermidiate  channels, $b$ is an impact parameter of the colliding hadrons. This equation is instrumental for the reconstruction of 
$h_{inel}(s,b)$\footnote{The function $h_{inel}(s,b)$ is
not well suited for the studies of asymptotics  due to  absence of the one-to-one correspondence between the functions $f(s,b)$ and $h_{inel}(s,b)$.}
from the elastic scattering data\footnote{Cf. e.g. \cite{ama} for an earlier analysis of $h_{inel}(s,b)$ and \cite{dremin} for the most recent one.}.

It should be noted that unitarity  implies an
existence   of the two scattering modes - absorptive and geometric ones.
Namely, the elastic scattering $S$-matrix (related to  the elastic scattering amplitude as $S(s,b)=1+2if(s,b)$)  
can be presented in the form
\[S(s,b)=\kappa(s,b)\exp[2i\delta(s,b)]\]
with the two real functions $\kappa(s,b)$ and $\delta(s,b)$. The function $\kappa$  
($0\leq \kappa \leq 1$)
 is an absorption factor\footnote{Its meaning is different in the reflection region, as it will be discussed further.},  its value $\kappa=0$ corresponds to a complete 
absorption. At the LHC energies the real part of the scattering amplitude would be neglected, i.e. we can perform replacement $f\to if$.
This assumption is  widely used .
Selection of the elastic scattering mode, namely, absorptive  or geometric   is governed then by the phase $\delta(s,b)$. 
The standard assumption is  $S(s,b)\to 0$ when the impact parameter $b$ is fixed and $s\to \infty$. It corresponds to 
a black disk limit  and the elastic scattering   is then completely absorptive, i.e. it is just a shadow of all the
inelastic processes. 

There is  another possibility, namely, the function $S(s,b)\to -1$ when $b$ is fixed 
and $s\to \infty$, i.e.  $\kappa \to 1$ and $\delta \to \pi/2$. This case corresponds to a pure geometric scattering \cite{reflect}. 
The principal point is that the phase is non-zero, i.e. $\delta$ is equal to $\pi/2$. 

We  discuss now 
the observable effects sensitive to the presence of the non--zero  phase. 
One of the useful quantities for that purposes is the ratio of elastic to total  impact--parameter dependent
 cross-sections, i.e.
\[
 {\cal R}(s,b)=\sigma_{el}(s,b)/\sigma_{tot}(s,b).
\]

The  impact--parameter dependent qauntities  $\sigma_{el}(s,b)$, $\sigma_{inel}(s,b)$ and $\sigma_{tot}(s,b)$ can be extracted from the
experimental data on the $d\sigma/dt$.
The function ${\cal R}(s,b)$ at the energies $s>s_0$ (the value of $s_0$ is determined by the solution of the equation $f(s,b=0)=1/2$) 
has two different forms depending on the value of impact parameter, namely at $b>R(s)$ this function has the form
\[{\cal R}(s,b)=[1-\kappa(s,b)]/2<1/2,\] while in the region $b<R(s)$ the presence of the non-zero phase (i.e. $\cos 2\delta=-1$)  changes this 
dependence to \[{\cal R}(s,b)=[1+\kappa(s,b)]/2>1/2.\] 
It means that the hadron scattering becomes predominantly  geometric one at small impact parameters and high energies $s>s_0$. 
Thus, the border of the region where reflective scattering is presented, is determined by the function $R(s)$. This is solution of the equation
\[f(s,b=R(s))=1/2\] at $s>s_0$.
The scattering in this case has a dominant absorptive contribution 
in the peripheral region of the impact parameters only.  The extraction from the experimental data of the function ${\cal R}(s,b)$ at
the finite energies can provide a hint on the asymtotics of the soft strong interactions, namely if the inequlity 
\[
{\cal R}(s,b)>1/2
\]
does  result from the experintal data analysis at small impact parameters and high values of the collision energy, 
one can arrive to conclusion that the black disk limit will be violated at the asymptotic energies
 and in this region the scattering would have a geometric origin since reflective scattering mode dominates \cite{reflect}.  
This would testify in favor of asymptotic limit  (\ref{rd}).
Of course, this claim is based on the assumed monotonic energy dependence of
the function ${\cal R}$ at high energies.

The  following asymptotic limit  takes place, namely,   ${\cal R}(s,b)\to 1$ when 
 impact parameter $b$ is fixed and $s\to\infty$ 
(i.e. geometric elastic scattering
saturates the total cross-section in this limit), 
while ${\cal R} (s,b)\to 0$ at fixed energy $s$
 and $b\to\infty$, i.e.  at large impact parameters elastic scattering cross-section decreases faster than the cross-sections of the all
inelastic processes, i.e. 
\[
 {\cal R}(s,b)={\bar{\cal R}}(s,b)/[1+{\bar{\cal R}}(s,b)],
\]
where
\[{\bar{\cal R}}(s,b)\equiv \sigma_{el}(s,b)/\sigma_{inel}(s,b)\to 0\]
 the following relation takes place
\[{\bar{\cal R}}(s,b)={f(s,b)}/{[1-f(s,b)]}\] at fixed $s$ and $b\to\infty$ and $f(s,b)\to 0$  in this limit since . 

Here the new ratio $ {\bar{\cal R}}(s,b)$ has been used.
It is clear
that \[{\bar {\cal R}}(s,b)\to \infty\] at fixed $b$ and $s\to\infty$ under transition to the non-zero phase  $\cos 2\delta=-1$. 
The function $\bar{\cal R}(s,b)$
is an another quantity sensitive to the presence of the reflective scattering mode and it determines the generalized reaction matrix dependence.
Namely, in the $U$--matrix approach for the unitarization (rational form of unitarization)
 the elastic scattering matrix element in the
impact parameter representation
is a linear fractional transform:
\begin{equation}
S(s,b)=\frac{1-U(s,b)}{1+U(s,b)}. \label{um}
\end{equation}
 $U(s,b)$ is the generalized reaction matrix, which is assumed to be 
input dynamical quantity. This relation (\ref{um}) is a one-to-one transform and can be easily inverted. 
The ratio ${\bar{\cal R}}(s,b)$ determines the function $U(s,b)$ , i.e.
 \[ U(s,b)={\bar{\cal R}}(s,b) \]
and this relation can be used as a tool for the reconstruction of  the function $U(s,b)$
from the experimental data. 

In the models the function $U(s,b)$  passes through unity with increasing energy providing that way a gradiual
transition to the geometric scattering mode. This transition implies  crossing zero value by the function $S(s,b)$ and aquiring the phase $\delta =\pi/2$
when the function $U(s,b)$ passes through unity. The  solution of the equation $U(s,b)=1$ separates the regions of absorptive and
geometric scattering and corresponds to the maximum value of $h_{inel}(s,b)=1/4$ since the derivative of $h_{inel}(s,b)$ has the form 
\[
 \frac{\partial h_{inel}(s,b)}{\partial b}=S(s,b)\frac{\partial f(s,b)}{\partial b} 
\]
and equals to zero at $U(s,b)=1$.  The derivative of the inelastic overlap function has the sign opposite to the sign of $\partial f(s,b)/\partial b$ in the region where $U(s,b)>1$ and 
the non-zero phase is,
therefore, responsible for the transformation of the central impact--parameter profile of the function $f(s,b)$ into a peripheral one of the inelastic overlap function $h_{inel}(s,b)$. The appearance
of the non-zero phase results (cf. Eq. \ref{um}) from transition  into the region where  $U(s,b)>1$. Since the function  $U(s,b)$ (its imaginary part in fact, but here we consider a 
pure imaginary case)
 gets contributions from the inelastic intermidiate channels only, its increasing behavior with energy can be associated with increasing contribution of the successive opening new
inelastic channels. This can be considered as  a dynamical origin for the relation $\cos 2\delta=-1$.

It should be noted that the derivative of the elastic overlap function has no sign-changing factor in front of $\partial f(s,b)/\partial b$, namely
\[
 \frac{\partial h_{el}(s,b)}{\partial b}=[1-S(s,b)]\frac{\partial f(s,b)}{\partial b}
\]
with  $1-S(s,b)$ being a non-negative at all values of $s$ and $b$.

Thus, the role of the non-zero phase in the high energy scattering is essential.
In the presence of the non--zero phase at the LHC energies the geometric  scattering dominates at small impact parameters while inelastic
processes are peripheral.  The albedo (coefficient of reflection) increases with energy 
 at $s>s_0$  \cite{reflect}. The factor $\kappa(s,b)$ plays the role of albedo at $s>s_0$ and $b<R(s)$ and hence should be considered as a reflective  factor rather than absorption factor 
when this region is taken into consideration.

This effect results in a number of consequences.
The perepherality of the inelastic amplitudes could be a dominating mechanism of the  ridge and double-ridge effects observed in  the two-particle correlation functions 
in proton-proton collisions \cite{coll,dri}. The  appearance of the non--zero phase at $s>s_0$ leading to the dominance of the geometric scattering, provides  change in the  slope (''the knee'') 
of the energy spectrum of the cosmic particles 
at the ground level  \cite{reflect}. 

At this point comments on the present experimental situation with elastic scattering at the LHC are to be done.  Most recent review of the data for elastic scattering with the analysis of the theoretical models has been given in \cite{dremin1}. As it was noted, the elastic scattering from the experimental viewpoint is consistent with BEL picture when the protons becomes blacker, edgier and larger \cite{henzi}. This is indicated by the analysis of the TOTEM 
data \cite{totem,totem1}. However, at the moment one cannot exclude the possibility of a gradual transition to the picture which name can be abbrevited as a REL one, i.e.
when the scattering region of the protons starts to become  reflective at the center and simultaneously edgier and larger at its peripherality. The TOTEM data do not provide a straightforward  clue for the asymptotic behavior, however the most recent luminosity--independent measurements at $\sqrt{s}=8$ TeV \cite{totem2} have confirmed steady 
increase of the ratio $\sigma_{el}/\sigma_{tot}$ providing the value $0.266\pm 0.006$ for this quantity. 

 The  processes at much higher collision energies can be observed in the cosmic rays.
 As it was noted in the beginning of the present paper, the cosmic rays data have significant error bars and therefore are not extremely conclusive. The highest--energy cosmic data 
 for the total inelastic cross-sections are provided now at the Pierre Auger Observatory \cite{auger}. Estimations on the base of these data \cite{dremin1}
 might be considered in favor of the REL scattering 
 picture since a decrease of the ratio  $\sigma_{inel}/\sigma_{tot}$ from 0.75 at
$\sqrt{s}=7$ TeV to 0.67 at $\sqrt{s}=57\pm 6$ TeV could take place. This result, however, should be taken with caution and  several reservations related to the low statistics
of the cosmic data points and model dependence of the extrapolations. 

Indirect information on the possible asymptotics can also be extracted  from the deep--elastic scattering  which is sensitive to the region of small impact
 parameters and from the studies of the correlations of elastic scattering process with the particle production processes \cite{del}.

 To get a more straightforward further information on which scattering mode takes place at asymptotic energies, namely, absorptive or geometric, one can use a full
reconstruction  of the impact--parameter dependent  scattering amplitude from the experimental data set. This set should include differential cross-section of the elastic 
scattering. 
Despite that such an analysis  assumes using the models for the phase of the scattering amplitude it has a full potential to extract the impact--parameter dependent quantities. 
We would like to point out that the recent analysis  \cite{martynov} implies that the amplitude $f(s,b)$ can reach and even
already crossed the black-disk limit  at $b=0$ and $\sqrt{s}=7$ TeV\footnote{It should be noted here that the value of $\mbox{Im} f(s,b=0)$ has increased
from $0.36$  (CERN ISR) to $0.492\pm 0.008$ (Tevatron)  and is very close to the black disk limit in this energy domain\cite{girom}.}. 
If it is so, it would testify  in favor of a gradual transition to the geometric elastic scattering starting at the LHC energies.

The model-independent analysis of the LHC data aiming to the reconstruction of the impact parameter amplitude $f(s,b)$
is desirable and would be instumental for the clarification 
of the hadron asymptotics. 
\section*{Acknowledgement}
We are grateful to E. Martynov  for 
 letting us know the  results of the impact parameter analysis of the elastic scattering.

\end{document}